\documentstyle[12pt]{article}

\begin{document}

\begin{center}
{\Large\bf A Possible Explanation of the Excess Jet Production at
Large $~E_T$}
\end{center}
\vskip 2cm

\begin{center}
{\bf Elliot Leader}\\
{\it Department of Physics\\
Birkbeck College, University of London\\
Malet Street, London WC1E 7HX, England\\
 E-mail: e.leader@physics.bbk.ac.uk}
\vskip 0.5cm
{\bf Dimiter B. Stamenov \\
{\it Institute for Nuclear Research and Nuclear Energy\\
Bulgarian Academy of Sciences\\
Blvd. Tsarigradsko chaussee 72, Sofia 1784, Bulgaria\\
E-mail:stamenov@inrne.acad.bg }}
\end{center}

\vskip 4.0cm
\begin{abstract}
We demonstrate that the significant discrepancy between
conventional QCD and the CDF inclusive jet production data could
be an indication that the behaviour of the Callan-Symanzik $~\beta$
function in QCD corresponds to it having a {\it finite fixed point}.
\end{abstract}
\vskip 0.5 cm
\newpage
\vskip 4mm

The inclusive jet production cross section in
$~p\bar{p}~\rightarrow jet + X~$ at $~\sqrt {s}=1.8~TeV~$, reported 
recently by the CDF Collaboration \cite{CDF}, is significantly higher
than the $~O(\alpha^3_{s})~$ QCD predictions of Ellis, Kunszt and
Soper \cite{EKS} for events with very high transverse energy 
($200~ \leq~E_T~\leq~440 GeV$), no matter which of a large set of
standard parton densities is utilised in the calculations. This has 
led to radical claims, for instance, that we are seeing evidence for    
the existence of constituents of quarks \cite{CDF, const}.\\

In this note we wish to suggest an unconventional, but much less radical,
solution to the problem. We remind the reader that the Callan-Symanzik 
$~\beta~$ function,
$~\beta(\alpha)~$, which controls the $~\mu^2~$- evolution of the
running coupling  $~\alpha_{s}(\mu^2)~$, in Perturbative QCD
is only calculable for small $~\alpha~$. From these calculations   
it follows \cite{GWP} that PQCD is asymptotically free  which
means that $~\alpha_{s}(\mu^2)~$ tends to zero as 
$~\mu^2~\rightarrow~\infty~$.
But this is true only if the $~\beta~$ function is {\it assumed} to 
be negative for all $~\alpha~$ up to $~O(1)~$ (solid curve in Fig. 1).\\

Some time ago one of us \cite{SiSt} investigated to what extent
the data on deep inelastic scattering \cite{DIS} for $~Q^2~\leq~500~GeV^2~$
and LEP data \cite {LEP} constrained the possibility of a so-called 
{\it fixed point} behaviour for QCD, in which $~\beta(\alpha)~$ could have 
one of the forms shown by the dotted and dashed curves in Fig. 1, corresponding
to a first or second order ultraviolet fixed point $~\alpha_0~$, respectively.
In the {\it fixed point} scenarios QCD is {\it not} asymptotically free 
(e.g. see the review \cite{Bur} and for the recent development
of the topic Ref. \cite{Pat})        
and $~\alpha_{s}(\mu^2)~\rightarrow~\alpha_0~ >~ 0~$ as 
$~\mu^2~\rightarrow~\infty~$. It was found that the data mentioned above are 
perfectly consistent with values
of $~\alpha_0 = 0.057~$ for a first order zero of $~\beta(\alpha)~$ and
$~0.029~\leq~\alpha_0~\leq~0.048~$ for the second order case. The essential
point relevant to the jet production cross section is the fact
that $~\alpha_{s}^{FP}(\mu^2)~$ runs more slowly than
$~\alpha_{s}(\mu^2)~$ does in QCD for large $~\mu^2~$, as shown in Fig. 2,
and will thus lead to cross sections which are enhanced  relative to the 
conventional QCD predictions as $~E_T~$ increases.\\

Let us now consider how we can estimate the difference between
the theoretical cross sections in the {\it fixed point} and conventional
versions of QCD. For fixed $~\sqrt {s}=1.8~TeV~$ and averaged
over pseudorapidity $~\eta~$ the theoretical cross section depend
upon $~E_T~$, the running coupling $~\alpha_s(\mu^2)~$ at scale
$~\mu~$ and the parton distributions $~f_{i}(x,\mu^2)~$ at the
same scale $~\mu~$. It is believed that $~\mu~$ should be of
order $~E_T~$ and the graphical comparison between the data and conventional
QCD shown in \cite{CDF} utilises $~\mu = E_{T}/2~$.\\

Now suppose that we have a conventional QCD calculation
$~\sigma_{n}~$ of the order $~(\alpha_{s})^{n}~$ contribution
to the cross section. The fixed point analogue $~\sigma^{FP}_{n}~$
will differ both on account of the different value of the running coupling 
and because the evolution of the parton densities is slightly different.
Based upon the studies in Ref. \cite{SiSt}, we believe the
latter to be very small effect.    

Hence in Born (leading order) approximation we expect to have

\begin{equation}
{d\sigma_{Born}^{FP}\over dE_{T}}(E_{T}, \mu, \alpha^{FP}_{s}(\mu^2))
\cong \left [{\alpha^{FP}_{s}(\mu^2)\over
\alpha_{s}(\mu^2)}\right ]^{2}
{d\sigma_{Born}\over dE_{T}}(E_{T}, \mu, \alpha_{s}(\mu^2))~.
\label{fpb}
\end{equation}\\

On other hand, as shown in \cite{EKS}, the Born cross section in conventional
QCD is essentially equal to the full (leading plus
next-to-leading) one at $~\mu = E_T~$:

\begin{equation}
{d\sigma_{Born}\over dE_{T}}(E_{T}, \alpha_{s}(E_{T}^2))
= {d\sigma_{Full}\over dE_{T}}(E_{T}, \alpha_{s}(E_{T}^2))~.
\label{BF}
\end{equation}\\

This result is obtained in \cite{EKS} for $~\eta=0~$ and cone size
$~R=0.6~$ at $~E_T = 100~GeV~$. For the kinematic range of interest
($R = 0.7~$ and $~\eta~$ averaged in the interval $~0.1~\leq~
\vert {\eta}\vert~\leq~0.7$) the value of $~\mu~$ at which Eq. (\ref{BF})
holds will be somewhat below $~\mu/E_{T} = 1~$ and its precise
value will depend weakly on $~E_T~$, but is essentially independent of  
$~E_T~$ in the range $~200~\leq~E_{T}~\leq~440~GeV~$ \cite{PrCom}.

Hence, using Eqs. (\ref{fpb}) and (\ref{BF}) we obtain the following
connection between $~\sigma_{Born}^{FP}~$ and $~\sigma_{Full}~$  
at $~\mu=E_T~$:

\begin{equation}
{d\sigma_{Born}^{FP}\over dE_{T}}(E_{T}, \alpha_{s}^{FP}(E_{T}^2))
= \left [{\alpha^{FP}_{s}(E_{T}^2)\over \alpha_{s}(E_{T}^2)}\right ]^{2}
{d\sigma_{Full}\over dE_{T}}(E_{T}, \alpha_{s}(E_{T}^2))~.
\label{muET}
\end{equation}\\

Strictly speaking Eq. (\ref{muET}) is an approximation.
However, bearing in mind that for $~\mu~$ in the range 
~$0.5\leq\mu/E_{T}<1~~~\sigma_{Full}~>~\sigma_{Full}(\mu=E_{T})$ 
the {\it fixed point}
enhancement over the conventional QCD cross section will be even
bigger than that arising from Eq. (\ref{muET}).\\

In order to illustrate the magnitude of the deviation of the theoretical
FP-QCD predictions from the conventional ones for the one jet production
cross section we will use for $~\alpha_{s}^{FP}(\mu^2)~$ in Eq. (\ref{muET})
the second order fixed point expression in LO approximation:

\begin{equation}
\alpha_s^{FP}(\mu^2) = \alpha_0 +
{\alpha_s(Q^2_0)-\alpha_0\over 1 + b~[\alpha_s(Q^2_0)-\alpha_0]~
ln~({\mu^2/Q_0^2})}~.
\label{aq2}
\end{equation}    

In (\ref{aq2}) the parameters $~\alpha_0~$ (the fixed point)
and $~b~$ are taken from Ref. \cite{SiSt} to be:
\begin{equation}
\alpha_0 = 0.040~,~~~~~~~b = 1~.
\label{par}
\end{equation}

The percentage difference between the theoretical FP-QCD cross sections
$~\sigma^{FP}~$ at $~\mu=E_T~$ (Eq. (\ref{muET}) above) and the conventional 
QCD predictions up to $~O(\alpha_{s}^3)~$
using the MRSD0\'~parton densities \cite{MRS} with scale $~\mu=E_{T}/2~$
is shown in Fig. 3 and compared with the percentage difference between the
CDF inclusive jet cross section data and the above mentioned conventional
QCD predictions.

It is seen that in the $~E_T~$ range $~15 - 200~GeV~$ the FP-QCD
predictions practically coincide with those of the conventional version
and are in good agreement with the data. However, for $~E_T~$
above $~200~GeV~$ the {\it fixed point} cross sections are higher than 
conventional QCD ones by up to $~20\%~$ at $~E_{T}=415~GeV~$ and are rather 
closer to the experimental points than the NLO QCD predictions.\\

In conclusion we suggest that the significant discrepancy between
conventional QCD and the CDF inclusive jet production data might be
an indication that the QCD $~\beta$ function corresponds to a fixed
point behaviour. If correct this fixed point behaviour will manifest   
itself clearly experimentally in that {\it all} large $~E_T~$ cross
sections will grow relative to the conventional QCD predictions once
$~E_{T}~\geq~200~GeV~$. We hope to provide detailed calculations
in the near future.\\

\vskip 4mm
{{\bf Acknowledgments}}
\vskip 3mm
The authors are grateful to Drs. D. Soper, M.G. Albrow, A. Bhatti
and A. Efremov  for helpful information and discussions.
One of us (E.L) would like to acknowledge the hospitality of the Institute
for Nuclear Research and Nuclear Energy of the Bulgarian Academy of
Sciences where this work was carried out.\\

This research was partly supported by the UK Royal Society, the UK PPARC
and by the Bulgarian Science Foundation under Contract \mbox{Ph 510.}\\

\end{document}